\documentclass[amsfonts,aps,draft,preprint]{revtex4}
\def\S2{\bar{S}}

\def\b{\beta}

\def\and{a_{n}^\dagger}

\def\sn2d{\Sn2^\dagger}

\def\({\left(}
\def\){\right)}
\def\<{\left\langle}
\def\>{\right\rangle}

\begin{document}

\title{On the entropy operator for the general $SU(1,1)$ TFD formulation}

\author{M. C. B Abdalla, A. L. Gadelha and Daniel L. Nedel}

\affiliation{Instituto de F\'{\i}sica Te\'{o}rica, Unesp, Pamplona
145, S\~{a}o Paulo, SP, 01405-900, Brazil }

\begin{abstract}
In this letter, an entropy operator for the general unitary $SU(1,1)$
TFD formulation is proposed and used to lead a bosonic system from
zero to finite temperature. Namely,
considering the closed bosonic string as the target system, the entropy
operator is used to construct the thermal vacuum.
The behaviour of such a state under the breve conjugation rules is
analized and it was shown that the breve conjugation does not affect the
thermal effects. From this thermal vacuum the thermal energy, the
entropy and the free energy of the closed bosonic string are calculated
and the apropriated thermal distribution for the system is found after
the free energy minimization.
\end{abstract}

\maketitle   

When the Thermo Field Dynamics (TFD) was proposed as an alternative
finite temperature quantum field theory \cite{tu}, an operator
called entropy was presented. Such a terminology comes from
the fact that its thermal expectation value is coincident with the
general expression for the entropy defined in quantum statistical
mechanics \cite{kubo}. Besides, it was shown that the entropy
operator can lead a system from zero to finite temperature.
Furthermore, it was identified as the time evolution controller
of systems, driving it through unitary inequivalent states
\cite{garvit}. In this context, the entropy operator was used to
study the vacuum structure of unstable particles \cite{fevit}
and quantum dissipation \cite{ceravi}, where in addition,
it appears as a dynamical variable leading to the conjecture that
these features reflect the irreversibility of time
evolution characteristic of dissipative motion \cite{ceravi}. Recently 
a TFD-like entropy operator was applied to show information loss in a
classical system \cite{blasone}, in connection with 't Hooft deterministic
quantum mechanics \cite{hooft1,hooft2}.

Returning to the thermal context, the entropy operator was 
recently applied to obtain the entropy of bosonic strings and the
thermal boundary states which is interpreted as $D$-branes at
finite temperature \cite{agv1,aev,agv3,agv4}. In these cases the very
form of the operator entropy was found considering that the finite
temperature system is obtained using a generator, $G$, satisfying the
following property
\begin{equation}
G1:\hspace{1cm}G\left( \theta \right)= G^{\dagger}\left( \theta
\right), \qquad \widetilde{G}\left( \theta \right)= - G\left( \theta
\right), \label{gprop0}
\end{equation}
meaning that, in a finite volume limit, the transformation is unitary
and preserves the tilde conjugation rules. In fact, this is the generator
presented in \cite{tu}. However, the TFD formalism can be generalized in
twofold. Both ways use as generator a linear combination of operators 
maintaining the nature of the thermal Bogoliubov transformation and form
a $SU(1,1)$ oscillator representation for bosons and $SU(2)$ for fermions. 
The differences between those two generalizations
refer to the behaviour of the total generator under tilde and
hermitian conjugations and can be put as follows:
\begin{eqnarray}
G2:\hspace{1cm}G\left( \theta \right)&\neq& G^{\dagger}\left( \theta
\right), \qquad \widetilde{G}\left( \theta \right)= - G\left( \theta
\right), \label{gprop1}
\\
G3:\hspace{1cm}G\left( \theta \right)&=& G^{\dagger}\left( \theta
\right), \qquad \widetilde{G}\left( \theta \right)\neq - G\left(
\theta \right). \label{gprop2}
\end{eqnarray}
In the first case the tilde conjugation rules are preserved, but the
transformation in a finite volume limit is non-unitary. This
construction was largely developed as one can see, for example, in Ref.
\cite{um1}. In the second case the transformation is unitary in a
finite volume limit, but the tilde conjugation rules are not
preserved. 

The unitary case, $G3$, was applied to bosonic strings and $D_{p}$-branes
in Ref. \cite{agv2} in order to investigate its statistical
properties. In this context, an entropy operator
was proposed considering its reduction to the original case
\cite{tu} when a suitable choice of the transformation parameters
is performed. As pointed out there, such a TFD generalization seems
to present an ambiguity in the thermal vacuum choice.

In a recent paper \cite{gfu}, the ambiguity related to the thermal
vacuum choice was shown to be just apparent and the general unitary $SU(1,1)$
TFD formulation was presented. Such a formulation considers a transformed
Tomita-Takesaki modular operator, since it does not commute with the
general generator of the thermal transformation \cite{elmume}. As a
consequence, the tilde conjugation rules were redefined in the
transformed space and named breve conjugation rules. The generalized
$SU(1,1)$ thermal vacuum is invariant under breve conjugation. However,
an analysis of the entropy operator in the general unitary $SU(1,1)$ TFD
formulation was not carried out. This analysis is the main point of this
letter. As it was said before the entropy operator can
be used to lead the system under study to finite temperature.
This property is explored here as a guidance to establish the appropriate
form of the entropy operator in the general $SU(1,1)$ TFD formulation.

The target system for our study is the closed bosonic string in the
light-cone gauge. To that end, the thermal vacuum obtained from the
entropy operator is requested to reproduce the statistical average of any
quantity $Q$  over an ensemble of one closed string states as follows
\cite{alvoso}, \cite{nos}:
\begin{eqnarray}
Z^{-1}(\beta)\;l_{s}\int dp^{+}\int d\lambda \;e^{\beta p^{+}} 
{\mbox Tr}\!\!\!\!&[&\!\!\!\!Q e^{-\frac{\beta}{p^{+}} H+2\pi i\lambda P}]
\nonumber
\\
&=&l_{s}\int dp^{+}\int d\lambda \int dp\; e^{\beta p^{+}}
\left\langle 0 (\b,\lambda, p^{+},p ) \left| Q
\right| 0 (\b,\lambda,p^{+},p )\right\rangle,
\nonumber
\\
\label{eqva}
\end{eqnarray}
where $Z(\beta)$ is the one string partition function
\begin{equation}
Z(\beta)=l_{s}\int dp^{+}\int d\lambda \;e^{\beta p^{+}}z_{lc}(\beta/p^{+})
\label{ospf}
\end{equation}
and ${z_{lc}(\beta/p^{+},\lambda)}$ the transverse light-cone one
\begin{equation}
z_{lc}(\beta/p^{+},\lambda)={\mbox Tr}[e^{-\frac{\beta}{p^{+}} H+2\pi i\lambda P}].
\end{equation}
In the above expressions, $l_{s}$ is the string fundamental length,
$p^{+}$ is the eigenvalue of the light-cone momentum operator,  $H$
is the light-cone hamiltonian, $P$ is the world-sheet momentum operator 
and $\lambda$ is a Lagrange multiplier. The above expectation value
takes into account the level matching condition of the closed
string Fock space and the thermal vacuum dependence on the Lagrange
multiplier $\lambda$ comes from the Bogoliubov transformation parameters.
As pointed out in \cite{nos} and shown later on this letter,
the explicit $\lambda$ dependence of these parameters
appears naturally when a free energy-like potential is minimized.
%
Finally, the integration over $p$ is related with the zero-mode 
contribution as it will be shown.

The comparison between the thermal vacuum obtained using the entropy
operator and that coming from the general Bogoliubov transformation
shows that the thermal states obtained by these
two different procedures differ by a phase factor. In this way,
the same thermal expectation value arises for these thermal states
and to that end, both states can be used to reproduce the statistical
average.

Consider the closed bosonic string. After its light-cone gauge
quantization, Fourier coefficients $\alpha ^{i}_{n}$ and $\beta ^{i}_{n}$,
with $i=1,...24$, for the left- and right-moving modes respectively, can be
redefined in order to obtain the creation and annihilation operators for
each mode $n$ in the different sectors. Namely,
\begin{eqnarray}
A_{n}^{i} &=&\frac{1}{\sqrt{n}}\alpha _{n}^{i},
\qquad A_{n}^{i\dagger}
=\frac{1}{\sqrt{n}}\alpha _{-n}^{i}, \\
B_{n}^{i} &=&\frac{1}{\sqrt{n}}\beta _{n}^{i},
\qquad B_{n}^{i\dagger}=\frac{ 1}{\sqrt{n}}\beta _{-n}^{i},
\end{eqnarray}
for every $n>0$. These redefined operators satisfy the oscillator-like
canonical commutation relations and the elements belonging to one sector
commute with those of the other. The fundamental state of the closed
bosonic string is defined by
\begin{equation}
A_{n}^{i}\left| 0\right\rangle =B_{n}^{i}\left| 0\right\rangle =0,
\end{equation}
where $\left| 0\right\rangle =\left| 0\right\rangle
_{\alpha }\left| 0\right\rangle _{\beta } $, as usual.

To construct the physical Fock space it is necessary to fix the
residual gauge symmetry generated by the world sheet momentum $P$.
This gauge fixing improves the level matching condition on a physical
state $\left |\Phi\>$:
\begin{equation}
P\left |\Phi\>= \sum_{n=1}^{\infty }\left( nA_{n}^{\dagger }\cdot
A_{n}- nB_{n}^{\dagger }\cdot B_{n}\right)\left |\Phi\>=0, 
\label{lmc}
\end{equation}
where the dot represents an Euclidean scalar product in the
transverse space.

Once the system is defined, to follow the TFD construction
one must now duplicate its degrees of freedom. In this way the
Fock space is now given by $\widehat{{\cal H}}={\cal H}\otimes
\widetilde{{\cal H}}$, where the tilde denotes a copy of the original
physical space. The vacuum of the doubled system is defined by
\begin{equation}
A_{n}^{i}\left. \left| 0\right\rangle \!\right\rangle
=B_{n}^{i}\left. \left| 0\right\rangle \!\right\rangle
=\widetilde{A}_{n}^{i}\left. \left| 0\right\rangle \!\right\rangle
=\widetilde{B}_{n}^{i}\left. \left| 0\right\rangle \!\right\rangle =0
\label{vdob}
\end{equation}
with $\left. \left| 0\right\rangle \!\right\rangle =\left. \left|
0\right\rangle \!\right\rangle _{\alpha }\left. \left| 0\right\rangle
\!\right\rangle _{\beta }$, in such a way that
\begin{eqnarray}
\left. \left| 0\right\rangle \!\right\rangle _{\alpha } &=&\left|
0\right\rangle _{\alpha }\otimes \widetilde{\left| 0\right\rangle }
_{\alpha }=\left| 0,\tilde{0}\right\rangle _{\alpha }, \\
\left. \left| 0\right\rangle \!\right\rangle _{\beta } &=&\left|
0\right\rangle _{\beta }\otimes \widetilde{\left| 0\right\rangle
}_{\beta }=\left|0,\tilde{0}\right\rangle _{\beta }.
\end{eqnarray}
The operator algebra of the duplicated system is now given by
\begin{eqnarray}
\left[ A_{n}^{i},A_{m}^{j\dagger }\right] &=&\left[
\widetilde{A}_{n}^{i},
\widetilde{A}_{m}^{j\dagger }\right] =\delta _{nm}\delta ^{ij}, \\
\left[ B_{n}^{i},B_{m}^{j\dagger }\right] &=&\left[
\widetilde{B}_{n}^{i}, \widetilde{B}_{m}^{j\dagger }\right] =\delta
_{nm}\delta ^{ij},
\end{eqnarray}
and all the other commutators are zero. Furthermore, according to the
TFD formulation the hamiltonian of the total (duplicated) system is
constructed in order to keep untouched the original dynamics of the
closed bosonic string. This operator can be written as
\begin{equation}
\widehat{H}=H-\widetilde{H},
\end{equation}
with
\begin{eqnarray}
H&=&\frac{P^{i} \cdot P^{i}}{2p^{+}}+
\frac{1}{p^{+}}\sum_{n=1}^{\infty }n\left( A_{n}^{\dagger }\cdot
A_{n}+B_{n}^{\dagger }\cdot B_{n}\right)-\frac{2}{p^{+}},
\nonumber
\\
\widetilde{H}&=&\frac{\widetilde{P}^{i}\cdot \widetilde{P}^{i}}{2p^{+}}+
\frac{1}{p^{+}}\sum_{n=1}^{\infty}
n\left( \widetilde{A}_{n}^{\dagger }\cdot \widetilde{A}_{n}+
\widetilde{B}_{n}^{\dagger }\cdot \widetilde{B}_{n}\right)-
\frac{2}{p^{+}},
\label{agas}
\end{eqnarray}
being the original and tilde system's light-cone hamiltonians,
respectively.

The next step in order reach the closed bosonic string at finite
temperature is to perform the thermal Bogoliubov transformation.
As it was mentioned, our interest is in a transformation which the
generator satisfies $G3$ given at (\ref{gprop2}). For the system
described above such a generator is
\begin{equation}
\mathbf{G}=G+\overline{G},
\end{equation}
where $G$ stands for the left-moving modes and $\overline{G}$ for the
right-moving modes, and explicitly for the left-moving modes only, it 
is given by
\begin{equation}
G = \sum_{n=1} \left[ \gamma _{1_{n}}\widetilde{A}_{n}^{\dagger }
\cdot A^{\dagger }_{n}
-\gamma_{2_{n}} A_{n}\cdot \widetilde{A}_{n} 
+\gamma
_{3_{n}}\left( A^{\dagger }_{n}\cdot A_{n} +
\widetilde{A}_{n} \cdot \widetilde{A}^{\dagger }_{n}\right)\right] ,
\label{ge}
\end{equation}
with the $\gamma$ coefficients given by
\begin{equation}
\gamma _{1_{n}} =\theta _{1_{n}}-i\theta
_{2_{n}}, \qquad \gamma_{2_{n}} =-\gamma
_{1_{n}}^{*},\qquad \gamma_{3_{n}} =\theta
_{3_{n}}.
\label{gammadef}
\end{equation}
The above $\theta$ parameters are functions of the Lagrange multiplier 
$\lambda$, $\b$, $p^{+}$ and $p$ as it will be shown in the last part of
this letter. Such a dependence was omitted on purpose as it will be in
the following, as well, for the sake of simplicity.

Still considering the left-moving modes, the thermal operators are
obtained by the following thermal Bogoliubov transformations
\begin{eqnarray}
\left(
\begin{array}{c}
A^{i}_{n}(\theta) \\
\breve{A}^{i \dagger}_{n}(\theta)
\end{array}
\right) &=&e^{-iG}\left(
\begin{array}{c}
A^{i}_{n} \\
\widetilde{A}^{i\dagger }_{n}
\end{array}
\right) e^{iG}={\mathbb B}_{n}\left(
\begin{array}{c}
A^{i}_{n} \\
\widetilde{A}^{i \dagger }_{n}
\end{array}
\right) , \label{tbt}
\\
\left(
\begin{array}{cc}
A^{i \dagger}_{n}(\theta) & -\breve{A}^{i}_{n}(\theta)
\end{array}
\right) &=&\left(
\begin{array}{cc}
A^{i \dagger }_{n} & -\widetilde{A}^{i}_{n}
\end{array}
\right) {\mathbb B}^{-1}_{n}, \label{tbti}
\end{eqnarray}
where the $SU(1,1)$ matrix transformation is given by
\begin{eqnarray}
{\mathbb B}_{n}=\left(
\begin{array}{cc}
u_{n} & v_{n} \\
v^{*}_{n} & u^{*}_{n}
\end{array}
\right) ,\qquad u_{n}u^{*}_{n}-v_{n}v^{*}_{n}=1, \label{tbm}
\end{eqnarray}
with elements
\begin{equation}
u_{n}=\cosh \left( i\Gamma_{n} \right) +\frac{\gamma
_{3_{n}}}{\Gamma_{n} } \mbox{senh}\left(i\Gamma_{n} \right), 
\qquad
v_{n}=\frac{\gamma _{1_{n}}}{\Gamma_{n} }\mbox{sinh}\left(
i\Gamma_{n} \right) , \label{uvexp}
\end{equation}
and $\Gamma_{n}$ is defined by the following relation
\begin{equation}
\Gamma ^{2}_{n}=\gamma _{1_{n}}\gamma _{2_{n}}+\gamma
_{3_{n}}^{2}.
\label{Gadef}
\end{equation}
The expressions (\ref{ge})-(\ref{Gadef}) remain the same for the
right-moving modes by replacing the $A$ by $B$ operators
and by taking the parameters $\theta$, $\gamma$, $\Gamma$
and the matrix element, with a bar.

Note that in the thermal Bogoliubov transformations given at
(\ref{tbt}) and (\ref{tbti}), the tilde symbol was replaced by the
breve symbol in the transformed operators. Such replacement realizes
the non preservation of the tilde conjugation rules under general
unitary thermal Bogoliubov transformation. The redefinition of such
rules was obtained by the thermal Bogoliubov transformation of the
Tomita-Takesaki modular operator, $J$, as presented in \cite{gfu}.
Namely,
\begin{equation}
J(\theta)=e^{-i\mathbf{G}}Je^{i\mathbf{G}}
=e^{-i\mathbf{G}}e^{-i\mathbf{G}^{*}}J
=Je^{i\mathbf{G}^{*}}e^{i\mathbf{G}}.
\label{jtran}
\end{equation}
The new conjugation rules in the transformed space, called breve,
can be written as follows: consider any operator $A\left(\theta \right)$
in the transformed space. Associated with this operator there is a breve
counterpart $\breve{A}\left(\theta \right)$. The map between both
is given by the following breve conjugation rules
\begin{eqnarray}
\left[ A\left(\theta \right)B\left(\theta \right)\right] \breve{^{}}
&=&\breve{A} \left(\theta \right)\breve{B}\left(\theta \right),
\label{breve1}
\\
\left[ c_{1}A\left(\theta \right)+ c_{2}B\left(\theta \right)\right]
\breve{^{}} &=&\left[
c_{1}^{*}\breve{A}\left(\theta\right)+c_{2}^{*}\breve{B}\left(\theta
\right)\right] ,
\label{breve2}
\\
\left[ A^{\dagger }\left(\theta \right)\right] \breve{^{}}
&=&\breve{A}^{\dagger }\left(\theta \right),
\label{breve3}
\\
\left[ \breve{A}\left(\theta\right)\right] \breve{^{}}
&=&A\left(\theta \right),
\label{breve4}
\end{eqnarray}
with $c_{1}$, $c_{2} \in \mathbb{C}$. Furthermore, the breve conjugation
rules leave invariant the thermal vacuum obtained from the thermal
Bogoliubov transformation with the generator given at (\ref{ge}):
\begin{equation}
\left[\left| 0\left(\theta\right)\right\rangle_{\mathbf{G}}\right]
\breve{^{}}
= J(\theta)\left| 0\left( \theta \right)
\right\rangle =
\left(e^{-i\mathbf{G}}e^{-i\mathbf{G}^{*}}J\right)e^{-i\mathbf{G}}
\left. \left| 0\right\rangle\!\right\rangle_{\mathbf{G}}
=e^{-i\mathbf{G}}\left. \left| 0\right\rangle\!\right\rangle
=\left| 0\left( \theta \right)
\right\rangle_{\mathbf{G}}.
\label{btvp}
\end{equation}
In this way one has,
\begin{eqnarray} 
\left[ \left| 0\left( \theta \right) \right\rangle_{\mathbf{G}}
\right] \breve{^{}} 
&=&\left| 0\left( \theta \right) \right\rangle_{\mathbf{G}} ,
\\
\left[{}_{\mathbf{G}}\left\langle 0\left(\theta\right)\right|\right]
\breve{^{}} 
&=&{}_{\mathbf{G}}\left\langle 0\left( \theta \right) \right|.
\label{rcbv}
\end{eqnarray}

It is interesting to note the possibility of rewriting the matrix
transformation in a more familiar form. In fact, 
if the polar decomposition is made:
$u_{n}=|u_{n}|e^{i\varphi}$,  $v=|v_{n}|e^{i\phi}$, after
the definitions
\begin{equation}
f_{n}=\frac{|v_{n}|^{2}}{|u_{n}|^{2}}, \qquad
\alpha_{n}=\frac{\log(\frac{v_{n}}{u_{n}})}{\log(f_{n})}
= \frac{1}{2}+i\frac{(\phi_{n}-\varphi_{n})}{\log(f_{n})},
\qquad
s_{n}=i\varphi_{n}
=\frac{1}{2}\log\left(\frac{u_{n}}{u_{n}^{*}}\right),
\label{sfap}
\end{equation}
the thermal Bogoliubov matrix is written in terms of the
parameters $s_{n}$, $\alpha_{n}$ and $f_{n}$ as follows
\begin{equation}
{\mathbb B_{n}}=\frac{1}{\sqrt{1-f_{n}}}\left(
\begin{array}{cc}
e^{s_{n}} & f_{n}^{\alpha_{n}}e^{s_{n}} \\
f_{n}^{1-\alpha_{n}}e^{-s_{n}} & e^{-s_{n}}
\end{array}
\right).
\end{equation}
The above matrix has the same formal expression that appears
in the non unitary general formulation as can be viewed in references
\cite{henume,hen}, for example. The similarity between the
two formulations was pointed out at references \cite{chume, igra}.
It is important to emphasize now that this similarity
is only formal. In the non unitary general formulation the matrix
elements are real, leaving the tilde rules invariant
by the thermal transformation. Here, in the unitary general
formulation, these elements are complex as manifested at
expressions (\ref{tbm}) and (\ref{sfap}). From the transformation
matrix point of view this fact leads to the necessity of redefining
the tilde conjugation rules making the breve conjugation rules
arise.

Let's now propose the general entropy operator for the system 
under study. It can be written as
\begin{equation}
\mathbf{K}=K+\overline{K} 
\label{kab}
\end{equation}
with the non bar and bar operators refering the to left- and
right-moving modes, respectively. For the left-moving modes one
has
\begin{equation}
K=-\sum_{n=1}\left[ A_{n}^{\dagger }\cdot A_{n}\log\left(
\frac{\gamma _{1_{n}}\gamma _{2_{n}}}{\Gamma
_{n}^{2}}\sinh^{2}\left( i\Gamma _{n}\right)\right) -A_{n}\cdot
A_{n}^{\dagger }\log \left( 1+\frac{\gamma _{1_{n}}\gamma
_{2_{n}}}{\Gamma_{n}^{2}}\sinh^{2}\left( i\Gamma
_{n}\right)\right)\right], \label{kag}
\end{equation}
and a similar expression for $\overline{K}$ replacing the $A$'s
operators by the $B$'s and considering the bar over the parameters.
The above expression is a direct generalization from that
presented in Ref. \cite{agv1}. The entropy
operator for the left- and right- moving modes can be written in
terms of the Bogoliubov matrix transformation elements in a compact
form. For the left-moving modes it is written as follows
\begin{equation}
K=-\sum_{n=1}\left[ A_{n}^{\dagger }\cdot A_{n}\log\left(f_{n}\right)
-g\log \left( |u_{n}|^{2}\right)\right],
\end{equation}
where $g=Tr(\delta^{ij})$ and with $u_{n}$ and $f_{n}$ given in
(\ref{uvexp}) and (\ref{sfap}), respectively.

The thermal vacuum is obtained from the entropy operator in a
similar way of that presented in Ref. \cite{tu} but with a suitable
introduction of the $\alpha$ parameter:
\begin{equation}
\left| 0\left( \theta \right) \right\rangle_{\mathbf{K}} =
e^{-\alpha K-\overline{\alpha}\overline{K}}
e^{\sum\limits_{n=1}\left[A_{n}^{\dagger }\cdot
\widetilde{A}^{\dagger}_{n}+B_{n}^{\dagger }\cdot
\widetilde{B}^{\dagger}_{n}\right]} \left. \left| 0\right\rangle
\!\right\rangle,
\end{equation}
with the $K$ and $\overline{K}$ operators defined at (\ref{kab}) and the
doubled vacuum $\left. \left| 0\right\rangle \! \right\rangle$, defined
in (\ref{vdob}). The label $\mathbf{K}$ in the thermal vacuum means only
the way employed to obtain this state. The $\alpha$ parameter was
considered the same for all modes.
By using the relations
\begin{equation}
e^{- \alpha K}\left. \left| 0\right\rangle\! \right\rangle =
\prod\limits_{n=1}\left( \frac{1}{|u_{n}|^{2} }\right) ^{\alpha g }
\left. \left| 0\right\rangle \!\right\rangle, \qquad e^{-\alpha
K}A_{n}^{\dagger \mu }e^{\alpha K}=
f^{\alpha}_{n} A_{n}^{\dagger \mu },
\end{equation}
with $\alpha_{n}$ and $f_{n}$ defined by (\ref{sfap}) and
$g=\mbox{Tr}[\delta^{ij}]=24$, the following condensed state expression
for the thermal vacuum is derived:
\begin{equation}
\left| 0\left( \theta \right) \right\rangle_{\mathbf{K}} = 
\prod\limits_{n=1}\left(
\frac{1}{|u_{n}|^{2} }\right)^{\alpha g}
\left(\frac{1}{|\overline{u}_{n}|^{2} }\right)^{\overline{\alpha}g} 
e^{f^{\alpha}_{n}
A_{n}^{\dagger \mu }\cdot \widetilde{A}_{n}^{\dagger \mu}
+\overline{f}^{\overline{\alpha}}_{n}B_{n}^{\dagger \mu }
\cdot \widetilde{B}_{n}^{\dagger \mu }}
\left. \left| 0\right\rangle \!\right\rangle,
\end{equation}
with its dual state given by
\begin{equation}
{}_{\mathbf{K}}\!\left\langle 0\left(\theta \right) \right|=
\left\langle\!\left\langle 0\right|\right. \prod\limits_{n=1}
\left(\frac{1}{|u_{n}|^{2} }\right)^{(1-\overline{\alpha}_{n}) g}
\left(\frac{1}{|\overline{u}_{n}|^{2} }\right)^{(1-\overline{\alpha}_{n})g}
e^{f^{1-\alpha_{n}}_{n}A_{n}^{\mu }\cdot\widetilde{A}_{n}^{\mu }
+\overline{f}^{1-\overline{\alpha}_{n}}_{n}B_{n}^{\mu }
\cdot \widetilde{B}_{n}^{\mu}},
\end{equation}
where $\alpha_{n}+\alpha_{n}^{*}=1$ (and an analogous one for
$\overline{\alpha}$) was used, resulting from $\alpha$ definition
at expression (\ref{sfap}).

Once the thermal vacuum is obtained from the general entropy
operator, its behaviour under breve conjugation rules can be studied.
At this point it should be noticed that this thermal vacuum can be
written in terms of the one coming from the use of the transformation
generator. In fact, considering
\begin{equation}
\left| 0\left( \theta \right) \right\rangle_{\mathbf{G}} = 
e^{-i\mathbf{G}}\left. \left| 0\right\rangle \!\right\rangle
=\prod\limits_{n}\left(
\frac{1}{u_{n}}\right)^{g}\left(\frac{1}{\overline{u}_{n}}\right)^{g}
e^{f^{\alpha}_{n}A_{n}^{\dagger\mu}\cdot \widetilde{A}_{n}^{\dagger\mu}
+\overline{f}^{\overline{\alpha}}B_{n}^{\dagger\mu}
\cdot \widetilde{B}_{n}^{\dagger\mu}}
\left. \left| 0\right\rangle \!\right\rangle,
\label{vacg}
\end{equation}
where label $\mathbf{G}$ sets out that it comes from the general generator,
one has, for each mode $n$ 
\begin{equation}
\left(\left| 0\left( \theta \right) \right\rangle_{K}\right)_{n}=
\left(e^{i\psi}
\left| 0\left( \theta \right) \right\rangle_{G}\right)_{n},
\end{equation}
where $e^{i\psi_{n}}\equiv \frac{u^{\alpha{*}}_{n}}{{u^{*}_{n}}^{\alpha}}$.
As the thermal vacua differ only by a phase, expectation values using 
$\left| 0\left( \theta \right) \right\rangle_{\mathbf{K}}$ or 
$\left| 0\left( \theta \right) \right\rangle_{\mathbf{G}}$
will be the same, independent of the choice.

Under breve conjugation rules (\ref{breve1})-(\ref{rcbv}), it follows that
\begin{equation}
\left[\left(\left|0\left(\theta\right)\right\rangle_{K}
\right)_{n}\right]\breve{^{}} =
\left(e^{-i\psi}
\left| 0\left( \theta \right) \right\rangle_{G}\right)_{n},
\end{equation}
since the thermal vacuum given at (\ref{vacg}) is invariant
under breve conjugation rules as shown at (\ref{btvp}).
Also the thermal vacuum is the same, but the phase.
Such a phase do not contribute to any thermal effect once that it
arises from expectation values in the thermal vacuum.

It is important to call the attention to the situation where one wants
to return to the TFD original formulation \cite{tu}. By choosing
$\alpha=1/2$ in the expression given at (\ref{sfap}), the
Bogoliubov matrix elements turn to be real and the phase $\psi$
vanishes. The thermal states obtained from the entropy operator and
from the transformation generator become exactly the same.

Let's now show how the statistical average over an ensemble of one closed
string states is reproduced by taking  expectation values on the thermal
vacuum above. To take into account only the physical states defined by
$\(\ref{lmc}\)$  it is necessary to improve the level matching condition
over the statistical average. In the TFD approach it can be done by
considering the shifted hamiltonian \cite{nos}, \cite{oji}:
\begin{equation}
H_{s}=H+\frac{ 2\pi i\lambda}{\beta}P
\end{equation}
where $H$, the original hamiltonian, is defined at (\ref{agas}),
$\lambda$ is a Lagrange multiplier and $P$ is given at (\ref{lmc}).
The dependence of the thermal vacuum on the Lagrange multiplier comes
from the Bogoliubov transformation parameters. This is achieved defining
first the free energy-like potential
\begin{equation}
{\cal F} = {\cal E} - \frac{1}{\beta}{\cal S},
\label{freed}
\end{equation}
where
\begin{eqnarray}
{\cal E} &\equiv& \left\langle 0 \left(\theta\right) 
\right| H_{s}\left|0\left(\theta\right)\right\rangle
\nonumber
\\
&=&g\frac{p^{2}}{2p^{+}}+
g\sum_{n}\left[n\left(|v_{n}|^{2}+|\overline{v}_{n}|^{2}|\right)
+\frac{2\pi i \lambda n}{\beta}\left(|v_{n}|^{2}-|\overline{v}_{n}|^{2}|
\right)\right]-\frac{2}{p^{+}},
\end{eqnarray}
is related to the thermal energy of the system, and
\begin{eqnarray}
{\cal S} &\equiv& \left\langle 0\left(\theta\right)
\right| K \left| 0\left( \theta \right) \right\rangle
\nonumber
\\
&=&- \sum_{n=1}\left[g|v_{n}|^{2}\log\left(f_{n}\right) -
g\log \left( |u_{n}|^{2}\right)
+g|\overline{v}_{n}|^{2}\log\left(\overline{f}_{n}\right) -
g\log \left( |\overline{u}_{n}|^{2}\right)\right]
\label{esdef}
\end{eqnarray}
with its entropy. By minimizing the free energy with respect to the
parameters, the following relations are found
\begin{eqnarray}
|v_{n}|^{2}&=&\frac{\gamma _{1_{n}}\gamma_{2_{n}}}
{\Gamma^2_{n}}\sinh^{2}\(i\Gamma_{n}\)
=\frac{1}{e^{n\left(\frac{\beta}{p^{+}}+i\lambda 2\pi\right)}-1},
\nonumber
\\
|\overline{v}_{n}|^{2}&=&
\frac{\overline{\gamma}_{1_{n}}\overline{\gamma}_{2_{n}}}
{\overline{\Gamma}^2_{n}}
\sinh^{2}\(i \overline{\Gamma}_{n}\)=
\frac{1}{e^{n\left(\frac{\beta}{p^{+}}-i\lambda 2 \pi\right)}-1},
\label{parmin}
\end{eqnarray}
where the bars specify the right-moving modes. The above expressions
show the dependence, mentioned before, of the transformation parameter
with the Lagrange multiplier $\lambda$.
Replacing the results (\ref{parmin}) in the expression for the potential
(\ref{freed}) one has
\begin{equation}
{\cal F}=-\frac{g}{\beta}
\ln\left\{\left[  e^{-\beta\frac{p^{2}}{2p^{+}}}e^{g^{-1}\frac{2\beta}{p^{+}}}
\prod_{n=1}\left|1-
e^{-n\left(\frac{\beta}{p^{+}}+2\pi i\lambda\right)}\right|^{-2}\right]
\right\}.
\label{frpex}
\end{equation}
Note that integrating the logarithm's argument over $p$ and using 
$g=\mbox{Tr}[\delta^{ij}]=24$, it results in
\begin{equation}
z_{lc}(\beta/p^{+},\lambda)
=\left[\left(\sqrt{\frac{2\pi p^{+}}{\beta}}\right)
e^{\frac{2\beta}{24p^{+}}}\prod_{n=1}
\left|1-e^{-n\left(\frac{\beta}{p^{+}}+2\pi i \lambda\right)}\right|^{-2}
\right]^{24},
\end{equation}
the transverse light-cone closed bosonic string partition function. Inserting
$z_{lc}(\beta/p^{+})$ in expression (\ref{ospf}) gives the one string
partition function
\begin{equation}
Z(\beta)=l_{s}\int dp^{+}\int d\lambda \;e^{\beta p^{+}} 
\left[\left(\sqrt{\frac{2\pi p^{+}}{\beta}}\right)
e^{\frac{2\beta}{24p^{+}}}\prod_{n=1}
\left|1-e^{-n\left(\frac{\beta}{p^{+}}+2\pi i \lambda\right)}\right|^{-2}
\right]^{24}.
\end{equation}
The expression for the one string partition function above is the same
obtained by using other approaches as can be found, for example,
in \cite{pol}.

Substituting (\ref{frpex}) into (\ref{eqva}) one finds
\begin{equation}
F(\beta)=-\frac{l_{s}}{\beta}\int dp^{+}\int d\lambda \;e^{\beta p^{+}} 
\ln\left[e^{-\beta\frac{p^{2}}{2p^{+}}}
e^{\frac{2\beta}{24p^{+}}}\prod_{n=1}
\left|1-e^{-n\left(\frac{\beta}{p^{+}}+2\pi i \lambda\right)}\right|^{-2}
\right]^{24}.
\end{equation} 
as the TFD response to the free energy of the closed bosonic string.
To finish, one can obtain the entropy of the closed bosonic string
substituting (\ref{parmin}) into (\ref{esdef}) and after in (\ref{eqva}).
The result can be written as
\begin{eqnarray}
S&=&l_{s}\int dp^{+}\int d\lambda \int dp\; e^{\beta p^{+}}
\nonumber
\\
&\times& \sum_{n=1}\left\{ \left[g_{n}\ln\left(1+\frac{n_{n}}{g_{n}}\right)
+n_{n}\ln\left(\frac{g_{n}}{n_{n}}+1\right)\right]
+\left[g_{n}\ln\left(1+\frac{\bar{n}_{n}}{g_{n}}\right)
+\bar{n}_{n}\ln\left(\frac{g_{n}}{\bar{n}_{n}}+1\right)\right]\right\},
\nonumber
\\
\end{eqnarray}
for $g_{n}=g=24$, and $n_{n}$, $\bar{n}_{n}$, the distributions
for left- and right-moving modes given at (\ref{parmin}).
The reason we write the entropy in such a way is to show
that this expression can be directly compared with the usual 
general expression for bosonic systems' entropy \cite{kubo},
with $g_{n}$ playing the r\^{o}le of degeneracy.

In this work the entropy operator of TFD was developed in order to construct
a unitary SU(1,1) thermal vacuum. By considering a Bogoliubov transformed
Tomita-Takesaki modular operator, it was shown that the SU(1,1) thermal
vacuum is invariant under redefined tilde conjugation rules. Also, 
the thermal vacuum constructed using the entropy operator differs from
the one achieved using the SU(1,1) unitary Bogoliubov generator by a phase.
Such a difference is not important when thermodynamical quantities are
derived.

The target system studied  was the closed bosonic string. It was shown
that expectation values in the  thermal vacuum reproduce the usual
thermodynamical results, obtained by evaluating the partition function in
the imaginary time formalism. However, the TFD approach can be used
further to give some new results in strings at finite temperature. 
As the partition function for the strings diverges at the Hagedorn
temperature \cite{Witten2}, the TFD  may be a very important tool to
understand the phase beyond the Hagedorn behaviour. In this case, the
thermal Fock space may be useful to identify new string's degrees of
freedom that can appear at hight temperature. In addiction, one can
use the formalism to go out of equilibrium and study dissipative processes
in string cosmology. Here, the  entropy operator will be very important,
playing the r\^{o}le of the non-unitary time evolution generator \cite{AMV}.

We would like to thank D. Z. Marchioro and Ademir E. Santana 
for usefull discussions.
M. C. B. A. was partially supported by the CNPq Grant 302019/2003-0,
A. L. G. and D. L. N. are supported by a FAPESP post-doc fellowship.


\begin{thebibliography}{99}

\bibitem{tu}
Y.~Takahasi and H.~Umezawa,
Collect.\ Phenom.\  {\bf 2} (1975) 55.

\bibitem{kubo}
M. Toda, R. Kubo and N. Saitao,
{\em Statistical Physics: Equilibrium Statistical Mechanics} 
(Springer-Verlag,1992).

\bibitem{garvit}
P.~Garbaczewski and G.~Vitiello,
Nuovo Cim.\ A {\bf 44} (1978) 108.

\bibitem{fevit}
S.~De Filippo and G.~Vitiello,
Lett.\ Nuovo Cim.\  {\bf 19} (1977) 92.

\bibitem{ceravi}
E.~Celeghini, M.~Rasetti and G.~Vitiello,
Annals Phys.\  {\bf 215} (1992) 156.

\bibitem{blasone}
M.~Blasone, P.~Jizba and G.~Vitiello,
Phys.\ Lett.\ A {\bf 287} (2001) 205
[arXiv:hep-th/0007138].

\bibitem{hooft1}
G.~'t Hooft,
arXiv:hep-th/0003005.

\bibitem{hooft2}
G.~'t Hooft,
Class.\ Quant.\ Grav.\  {\bf 16} (1999) 3263
[arXiv:gr-qc/9903084].

\bibitem{agv1}
M.~C.~B.~Abdalla, A.~L.~Gadelha and I.~V.~Vancea,
Phys.\ Rev.\ D {\bf 64} (2001) 086005
[arXiv:hep-th/0104068].

\bibitem{aev}
M.~C.~B.~Abdalla, E.~L.~Graca and I.~V.~Vancea,
Phys.\ Lett.\ B {\bf 536} (2002) 114
[arXiv:hep-th/0201243].

\bibitem{agv3}
M.~C.~B.~Abdalla, A.~L.~Gadelha and I.~V.~Vancea,
Int.\ J.\ Mod.\ Phys.\ A {\bf 18} (2003) 2109
[arXiv:hep-th/0301249].

\bibitem{agv4}
M.~C.~B.~Abdalla, A.~L.~Gadelha and I.~V.~Vancea,
arXiv:hep-th/0308114.

\bibitem{um1}
H. Umezawa, {\em Advanced Field Theory: Micro, Macro and Thermal Field }
(A\-me\-ri\-can Institute of Physics, 1993).

\bibitem{agv2}
M.~C.~B.~Abdalla, A.~L.~Gadelha and I.~V.~Vancea,
Phys.\ Rev.\ D {\bf 66} (2002) 065005
[arXiv:hep-th/0203222].

\bibitem{gfu}
M.~C.~B.~Abdalla and A.~L.~Gadelha,
Phys.\ Lett.\ A {\bf 322} (2004) 31
[arXiv:hep-th/0309254].

\bibitem{elmume}
P.~Elmfors and H.~Umezawa,
Physica A \textbf{202} (1994) 577.

\bibitem{alvoso}
E.~Alvarez and M.~A.~R.~Osorio,
Phys.\ Rev.\ D {\bf 36} (1987) 1175.

\bibitem{nos}
Daniel L.~Nedel, M.~C.~B.~Abdalla and A.~L.~Gadelha,
Phys. Lett. B {\bf 598} (2004) 121
[arXiv:hep-th/0405258].

\bibitem{henume}
P.~A.~Henning and H.~Umezawa,
Nucl.\ Phys.\ B {\bf 417} (1994) 463.

\bibitem{hen}
P.~A.~Henning,
Phys.\ Rept.\  {\bf 253} (1995) 235.

\bibitem{chume} H. Chu and H. Umezawa,
Int.\ J.\ Mod.\ Phys.\ A{\bf 9} (1994) 2363.

\bibitem{igra}
M.~C.~B.~Abdalla, A.~L.~Gadelha and I.~V.~Vancea,
Nucl.\ Phys.\ Proc.\ Suppl.\  {\bf 127} (2004) 92.

\bibitem{oji}
I.~Ojima,
Annals Phys.\  {\bf 137} (1981) 1.

\bibitem{pol}
J.~Polchinski.
``String Theory'', Vol. I,
(Cambridge University Press, Cambridge, 1998).

\bibitem{Witten2}
J.~J.~Atick and E.~Witten,
Nucl.\ Phys.\ B {\bf 310} (1988) 291.

\bibitem{AMV}
E.~Alfinito, R.~Manka and G.~Vitiello,
Class.\ Quant.\ Grav.\  {\bf 17} (2000) 93
[arXiv:gr-qc/9904027].
 
\end{thebibliography}
\end{document}